\documentclass[aip,superscriptaddress,twocolumn,tightenlines,10pt,amssymb,amsmath, showpacs]{revtex4-1}
\usepackage{graphicx}
\usepackage{subfigure}
\usepackage{natbib}
\usepackage{longtable}
\usepackage{epstopdf}
\usepackage{tabularx}
\usepackage{graphicx}
\usepackage{hyperref}
\usepackage{xcolor}
\usepackage[latin1]{inputenc}
\usepackage{amsmath}
\usepackage{amsfonts}
\usepackage{amssymb}
\usepackage{hyperref}
\hypersetup{
  colorlinks   = true, 
  urlcolor     = blue, 
  linkcolor    = red, 
  citecolor   = blue 
}
\begin{document}

\title{Surface Optical and Bulk Acoustic Phonons in the Topological Insulator, Bi$_2$Se$_2$Te}
\author{Uditendu Mukhopadhyay \cite{contrib}}
\author{Dipanjan Chaudhuri \cite{contrib}}
\author{Jit Sarkar}
\author{Sourabh Singh}
\author{Radha Krishna Gopal}
\author{Sandeep Tammu}
\affiliation{Indian Institute of Science Education and Research (IISER) Kolkata,
Mohanpur  741246, Nadia, West Bengal, India.}

\author{Prashanth C. Upadhya}
\affiliation{Laboratory for Electro-Optics Systems, Indian Space Research
Organization, Bangalore 560058, India.}

\author{Chiranjib Mitra \cite{email}}
\affiliation{Indian Institute of Science Education and Research (IISER) Kolkata,
Mohanpur  741246, Nadia, West Bengal, India.}
\email{chiranjib@iiserkol.ac.in}

\date{\today}

\begin{abstract}
We explore the phonon dynamics in thin films of the topological insulator material Bi$_2$Se$_2$Te using ultrafast pump-probe spectroscopy. The time resolved differential reflectivity in these films exhibit fast and slow oscillations. We have given a careful analysis of variation of phonon frequency as a function of film thickness, which we  attribute to the existence of standing acoustic modes. However, no variation in the frequency of the optical phonon modes was found with film thickness. This indicates that the optical phonons intrinsically belong to the surface of topological insulators. The controllability of acoustic phonons by way of varying the film thickness will have tremendous implications in the application of these materials in low power spintronic device operating at room temperature. 
\end{abstract}


\maketitle
The newly discovered 3D topological insulators (TIs) represent a class of exotic quantum state of matter with an insulating bulk and conducting surface states \cite{kane2004physics}$^,$\cite{fu2007topological}. These conducting surface states exhibit a relativistic energy-momentum dispersion owing to the nontrivial Z$_2$ topology of the bulk electronic wave-function\cite{hasan2010colloquium}. As a consequence, these surface states form a Dirac cone between bulk valence band and bulk conduction band which has been confirmed by angle resolved photo-emission spectroscopy (ARPES) studies \cite{hsieh2008topological}$^,$\cite{xia2009observation}. Additionally, the surface states are ``topologically'' protected from back-scattering. Helical spin texture due to high spin-orbit interaction, a $\pi$ Berry phase and the presence of odd number of Dirac cones are responsibe for this topological protection. Furthermore, the constraint of time reversal symmetry ensures that the surface electrons are not scattered from non-magnetic impurities and defect states. Apart from these, TI materials also exhibit interesting magneto-optical properties which arise due to the presence of axionic term \citep{aguilar2012terahertz}. These exciting properties have made TIs a subject of intense research both theoretically as well as experimentally over the past few years. TIs also promise important practical application in quantum computation and low  power spintronic device applications \cite{moore2010birth}. \\~\\

The family of Bismuth based compounds including Selenides and Tellurides are well established examples of TI materials\citep{zhang2009topological}. Previously, ultrafast pump-probe spectroscopy has been employed to study the phonon and carrier dynamics in Bi$_2$Se$_3$ (BS) \cite{kumar2011spatially}$^,$\cite{glinka}, Bi$_2$Te$_3$ (BT)\cite{zhang}$^,$\cite{misochko2015polarization}$^,$\cite{weiss2015ultrafast}, Sb$_2$Te$_3$ \cite{:/content/aip/journal/jap/117/14/10.1063/1.4917384} etc. The differential reflectivity measured by pump-probe spectroscopy revealed fast and slow oscillations which have been attributed to surface optical and bulk acoustic phonons respectively \cite{qi}. However these claims have not been conclusively substantiated. In this letter, we reinvestigate these claims through similar single coloured, optical pump-probe spectroscopy in reflection geometry in TI films of different  thicknesses. Most of the existing research have been carried out on BS and BT systems which have dominant bulk carriers owing to defect states that overwhelm the surface conduction \cite{ando}. In our work, we have used Bi$_2$Se$_2$Te (BST) thin films, which exhibit lower bulk conduction owing to reduced defect states \cite{bao}. Moreover, BST is also predicted to have a large spin texture giving rise to non-trivial Berry phase that prevents back-scattering \cite{bao}$^,$\cite{wang}. To support this claim, variation of the resistance with temperature has been shown in the supplementary material \cite{Sup}, clearly displaying the bulk insulating feature. We observed both fast and slow oscillations in BST thin films.\\~\\

BST thin films used in our experiments have been deposited on Silicon (100) substrates using pulsed laser deposition (PLD) technique. Samples of thickness 50 nm, 100 nm and 300 nm were prepared in identical conditions by varying the number of laser pulses. Further details of sample preparation have been reported elsewhere \cite{gopal}. Furthermore, detailed information on the characterisation of these thin films are provided in the supplementary material \cite{Sup}. An ultrafast femtosecond pulsed laser (MANTIS from Coherent Inc. USA) was used to generate Gaussian pump and probe pulses of 780 nm center wavelength and 100 fs pulse-width with a repetition rate of 80 MHz. The pump beam was modulated at 1.8 KHz using a mechanical chopper and pump induced change in the reflectivity of probe beam was measured using a lock-in amplifier as a function of pump-probe delay \cite{Udit}. The temporal resolution in our measurement was 66 fs (from 0 to 15 ps delay) and 333 fs for the rest of the scan. Polarization of pump and probe pulse were made perpendicular to each other to reduce pump scattering. All measurements reported in this letter were performed at room temperature.\\

\begin{figure}
\begin{center}
\includegraphics[scale=0.35]{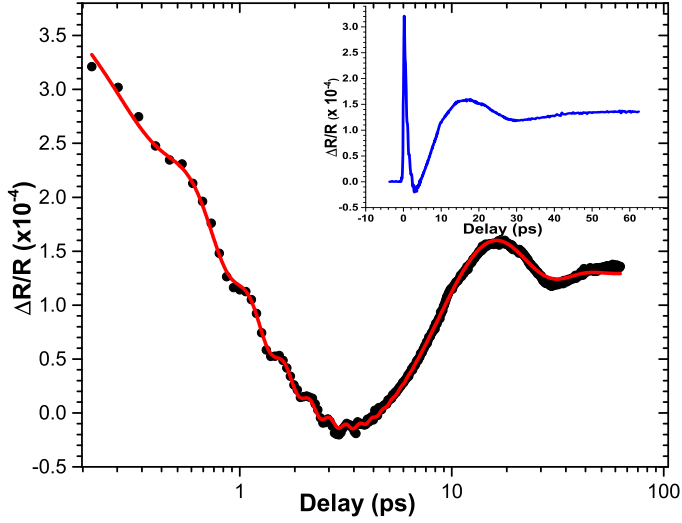}
\caption{Differential reflectivity of 100 nm thick BST film measured at room temperature is plotted (scattered plot) in semi-logarithmic scale. The solid line is the fitted curve. Inset shows the same data plotted in linear scale capturing the zero delay.}\label{fig:fulloscillation}
\end{center}
\end{figure}

~\\Time resolved differential reflectivity in BST film is shown in fig.\ref{fig:fulloscillation} (inset). A fast rise time indicates rapid transition of electrons from valence to conduction band. The semi-logarithmic plot of the data shows short lived rapid oscillations superimposed on rather slower oscillations that lasts for a longer duration. The fast oscillations last for around 10 ps, whereas, the slower one persists for around 100 ps. Measurements were carried for longer pump-probe delay (800 ps) but no significant modulation was observed after 100-120 ps.\\~\\

The differential reflectivity in fig.\ref{fig:fulloscillation} is fitted with double decay function to capture the overall carrier relaxation dynamics. The damped oscillation terms were added to fitting equation to extract the characteristic frequencies along with the appropriate decay times \citep{cheng}. The fitting function is thus given by,
\begin{eqnarray}\label{eq:fit}
\frac{\Delta R}{R} &=& A_1 \exp \left\lbrace -(t-t_0)/t_1\right\rbrace + A_2 \exp \left\lbrace -(t-t_0)/t_2\right\rbrace \nonumber\\
 &+& B_1 \exp \left\lbrace -(t-t_0)/t_3\right\rbrace sin (2\pi f_1 + \phi_1) \nonumber\\
 &+& B_2 \exp \left\lbrace -(t-t_0)/t_4\right\rbrace sin (2\pi f_2 + \phi_2) +C 
\end{eqnarray}
~\\

\begin{figure}
\begin{center}
\includegraphics[width =0.5 \textwidth]{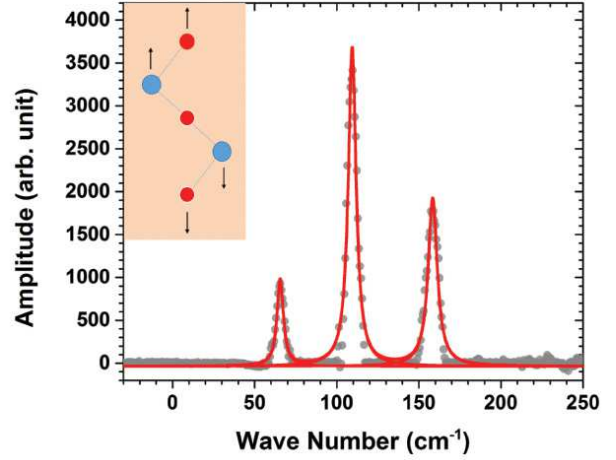}
\caption{Raman scattering data (scattered plot) and its Lorentzian fit (solid line) of the BST thin film. A$^1_{1g}$ Raman peak is at 65.6 cm$^{-1}$ (1.97 THz). The other modes E$_{1g}$ and A$^2_{1g}$ are at 109.67 cm$^{-1}$ and 158.74 cm$^{-1}$ respectively. Inset shows A$^1_{1g}$ optical phonon mode vibration. Larger circles represent Bi atoms wheras smaller circles represent Se and Te atoms.}\label{fig:raman}
\end{center}
\end{figure}

From the fit, the extracted frequencies of fast and slow oscillation are found to be 1.93 THz and 0.03 THz respectively. These fast and slow oscillations are attributed to surface optical and bulk acoustic phonon modes respectively \cite{qi}. To confirm, Raman scattering spectra of these samples were taken using blue laser (488 nm) with a  power of 6 mW and a magnification of 50$\times$. The charistic peaks are observed at 65.67 cm$^{-1}$, 109.67 cm$^{-1}$ and 158.74 cm$^{-1}$ (see fig.\ref{fig:raman}). Previous studies \cite{richter} attributed these peaks to the A$^1_{1g}$, E$_{1g}$ and A$^2_{1g}$ Raman modes in BST (with hexagonal crystal structure) respectively. Frequency corresponding to the A$^1_{1g}$ Raman mode (1.97 THz) is in close agreement to the surface optical phonon mode observed in our pump probe measurements. The inset of fig.\ref{fig:raman} shows the vibration corresponding to the A$^1_{1g}$ Raman mode. However, for the detection of lower symmetry E$_g$ modes anisotropic polarization dependent measurements are necessary \cite{misochko2015polarization, :/content/aip/journal/jap/117/14/10.1063/1.4917384}. The A$^2_{1g}$ coherent phonons in Bismuth based systems have been reported to decay faster than A$^1_{1g}$ phonons and thus making them difficult to detect \cite{Norimatsu201358, PhysRevB.88.064307}.\\

\begin{figure}
\begin{center}
\includegraphics[scale=0.3]{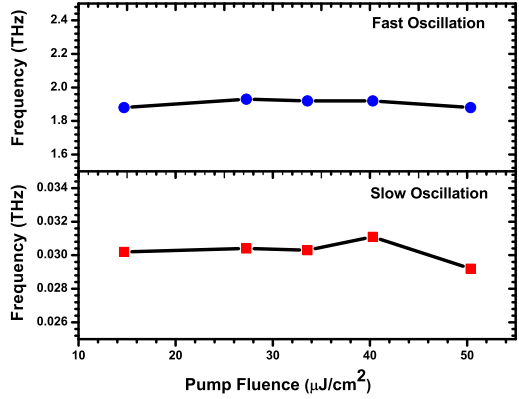}
\caption{Dependence of pump fluence on fast (top panel) and slow oscillation (bottom panel) in BST thin film with a thickness of 100nm.}\label{fig:fluence}
\end{center}
\end{figure}

~\\To understand the excitation density dependence, the pump fluence was varied from 15 $\mu$J/cm$^2$ to 50 $\mu$J/cm$^2$ while maintaining a constant probe fluence of 1.25 $\mu$J/cm$^2$. Frequencies of both the fast and slow oscillations do not show any significant variation (see fig.\ref{fig:fluence}). This indirectly supports the hypotheses that the two oscillations represent the phonon dynamics as change in  the pump fluence will only change the carrier density in the conduction band and is not expected to have any effect on either of the phonon modes. However, we note that the amplitude of coherent oscillations changes almost linearly with fluence. \\

\begin{table*}[t]
\begin{center}
\bgroup
\def\arraystretch{1.5}
\begin{tabular}{c c c c c c c}
\hline
\hline\\
Sample &~& \multicolumn{2}{c}{Fast Oscillation} &~& \multicolumn{2}{c}{Slow Oscillation} \\ 
 Thickness &~& Frequency  & Damping  &~& Frequency  & Damping\\  
 (nm) &~& (THz) & (ps) &~& (THz) & (ps)\\   
 \hline
 \hline
 50 &~& 1.960 $\pm$ 0.008 & 1.41 $\pm$ 0.03 &~& 0.0330 $\pm$ 0.0008 & 4.79 $\pm$ 0.21\\
 \hline
 100 &~& 1.914 $\pm$ 0.022 & 1.40 $\pm$ 0.05 &~& 0.0299 $\pm$ 0.0010 & 8.42 $\pm$ 0.59\\
 \hline
 300 &~& 1.917 $\pm$ 0.049 & 1.33 $\pm$ 0.03 &~& 0.0149 $\pm$ 0.0010 & 25.34 $\pm$ 1.03\\
 \hline
 \hline
\end{tabular}
\caption{Observed frequency and damping of the fast and slow oscillations}
\egroup
\end{center}
\end{table*}
~\\

\begin{figure}
\begin{center}
\includegraphics[scale=0.3]{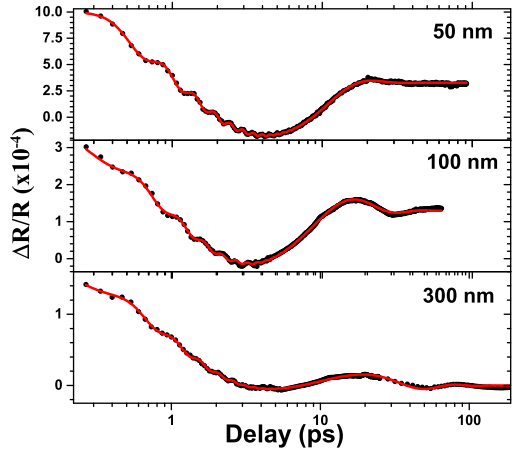}
\caption{Fitting (solid line) of differential reflectivity data (scattered plot) for BST samples of thickness (a) 50 nm, (b) 100 nm, (c) 300 nm.}\label{fig:thickness}
\end{center}
\end{figure}

\begin{figure}
\begin{center}
\includegraphics[scale=0.3]{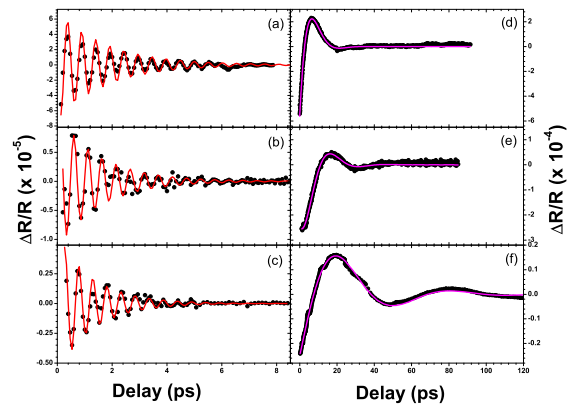}
\caption{Fast and slow oscillations for different samples as extracted from fig.\ref{fig:thickness}. The dotted lines represent experimental data and solid lines represent the corresponding fit.  (a)-(c): Fast oscillations in 50 nm, 100 nm and 300 nm films respectively. (d)-(f): Slow oscillations in 50 nm, 100 nm and 300 nm films respectively.}\label{fig:full}
\end{center}
\end{figure}

~\\
The same experiment was then repeated on identically prepared BST films of thicknesses 50 nm, 100 nm and 300 nm. Any change in the thickness is expected to alter the boundary conditions for the acoustic phonon oscillations. In fig.\ref{fig:thickness}, the pump-probe signal for the three different samples are shown along with their fitting. The magnitude of differential reflectivity is significantly reduced with the increasing thickness. To illustrate this further, the fast and the slow oscillations for all three samples are separately plotted in fig.\ref{fig:full} along with the fit. From this it can be easily concluded that both the frequency of oscillation and the damping factor for the fast oscillations remain fairly constant with varying thickness. However, this fast oscillation becomes more prominent in the thinner sample. Here, the damping factor of fast oscillations indicates the timescale of phonon phonon interactions \cite{qi}.\\
~\\
On the other hand, the slow oscillations are significantly affected by the change in thickness. The frequency of these oscillation decreases with increasing thickness. This can be explained by assuming that the slow oscillations are a result of interference between multiple reflection of the acoustic wave from the vacuum-sample and sample-substrate boundaries. The result of this interference is a standing wave like modulation which is expected to change with thickness as the boundary conditions are being effectively altered. This result is qualitatively similar to the normal mode oscillation of a string with fixed ends, where the frequency of oscillation is inversely proportional to the length of the string.\\~\\

Thickness of the sample also shows a pronounced effect on the damping of the slow oscillations. The damping can be physically attributed to the strain owing to the lattice mismatch between the sample and the substrate. Whereas hexagonal quintuple BST layers have a lattice constant of 4.218 \AA \cite{lat} and Si(100) has a diamond cubic lattice structure with lattice spacing of 5.431 \AA. Because of this lattice mismatch, there is a mechanical strain on the BST layers deposited. However, with increasing thickness, the BST films are allowed more degrees of freedom, which helps in relaxing this strain, resulting in reduced damping in the slow oscillations. From the fitting of fig.\ref{fig:thickness}, the fast and and slow oscillations are extracted and plotted separately in fig.\ref{fig:full}. The slow oscillations clearly demonstrate this reduction in damping with increasing thickness. Amplitude of the slow oscillations also decreases with increasing thickness. As mentioned above, the damping of the faster oscillations remains constant for different samples indicating that these are indeed characteristic of a surface phenomenon without any bulk contribution. Therefore, the optical phonons corresponding to the A$^1_{1g}$ mode, intrinsically belong to the surface of the TI material.\\
 
The ultrafast phenomena that induces optical phonon oscillations at the surface and acoustic phonon oscillations in the bulk can be understood in the following way. The photo excited carriers experience a non-linear ponderomotive force at the surface \citep{PhysRevB.88.064307} and excite the cold lattice to oscillate coherently at the optical-phonon frequency (1.97 THz). The fast de-phasing of these oscillations are attributed to the existence of surface states \citep{weiss2015ultrafast}. In the bulk, however, carriers relax through electron-acoustic phonon interaction process which is slower. Owing to the fact that the bulk is insulating in these materials (BST) and the Fermi level resides in the bulk band gap, ultrafast dynamics of the carriers originating from the bulk is very different from that of surface \cite{gopal1}, unlike in their parent compounds (BT and BS). \\~\\

From the perspective of device applications, though the surface electronic states are topologically protected against scattering from non-magnetic defects, acoustic phonons are still an impediment to the electron flow in the bulk. This is because, electron-phonon scattering will always be there at room temperature. For good device applications, in principle, one needs to lower the temperature to minimise the electron-phonon coupling. Moreover, the controllability of frequency of acoustic phonons by varying film thickness has a bearing on improving the low power spintronic device application and in certain cases, even without having to cool the system. \\

In conclusion, the experiment captures the optical and acoustic phonon modes through time resolved differential reflectivity measurements in BST thin films. Short lived fast oscillations, which are superimposed on the pump-probe signal are attributed to the A$^1_{1g}$ optical phonon mode. The slower oscillations correspond to modulation due to interference of bulk acoustic phonon modes. By studying films of different thicknesses, we have been able to conclusively establish that optical phonon modes are indeed a surface phenomenon with no contribution form the bulk. On the other hand, the slow oscillations corresponding to the acoustic phonon modes are significantly affected by film thickness and this has been qualitatively explained in the light of dynamics of mechanical standing waves in elastic media. Our results will have significant impact on the application of topological insulators in spintronic device and topological quantum computation.\\
~\\
The Authors would like to thank Ministry of Human Resource Development (MHRD) for funding. JS and SS would like to thank U.G.C. for financial assistance. CM would like to thank Anjan Barman for initial experiments though not successful but nevertheless educative.
%

\begin{thebibliography}{21}
\bibitem[$\dagger$]{contrib} These two authors have equal contribution
\end{thebibliography}

\end{document}